\documentclass[prd,twocolumn,showpacs,preprintnumbers,amsmath,amssymb,superscriptaddress,floatfix,nofootinbib]{revtex4-2}

\usepackage{graphicx}
\usepackage{bm}
\usepackage{amsmath}
\usepackage{amsfonts}
\usepackage{amssymb}
\usepackage{color}
\usepackage{multirow}
\usepackage{subfigure}
\usepackage{txfonts}
\usepackage[colorlinks, citecolor=blue,anchorcolor=red,menucolor=red, linkcolor=red,filecolor=red,runcolor=red,urlcolor=blue,frenchlinks=red]{hyperref}

\begin{document}

\title{Probing the isospin structure and low-lying resonances in $\Lambda_c^+ \to n\bar{K}^0  \pi^+$ decays}

\date{\today}
\author{Meng-Yuan Li}
\affiliation{School of Physics, Zhengzhou
	University, Zhengzhou 450001, People’s Republic of China}

\author{Guan-Ying Wang}\email{wangguanying@henu.edu.cn}
\affiliation{School of Physics and Electronics, Henan University, Kaifeng 475004, People’s Republic of China}

\author{Neng-Chang Wei}\email{weinengchang@htu.edu.cn}
\affiliation{School of Physics, Henan Normal University, Xinxiang 453007, People’s Republic of China}
 
\author{De-Min Li}\email{lidm@zzu.edu.cn}
\affiliation{School of Physics, Zhengzhou
	University, Zhengzhou 450001, China}

\author{En Wang}\email{wangen@zzu.edu.cn}
\affiliation{School of Physics, Zhengzhou
	University, Zhengzhou 450001, People’s Republic of China}

\begin{abstract}
The Cabibbo-favored decay $\Lambda_c^+ \to n \bar{K}^0\pi^+$ offers a unique window to explore unresolved puzzles in the low-energy baryon spectroscopy and the isospin dynamics of the $\bar{K}N$ system. Recent experimental results present a, for now, contradiction: LHCb and Belle analyses of $\Lambda_c^+ \to p K^-\pi^+$ suggest the $pK^-$ ($I=0$) component dominates, while the Beijing Spectrometer III (BESIII) hints at significant contributions from both isospin $0$ and $1$ in the $n\bar{K}^0$ system of $\Lambda_c^+ \to n K_S^0 \pi^+$. Furthermore, the measured branching fraction of $\Lambda_c^+ \to n K_S^0 \pi^+$ exceeds SU(3) symmetry predictions by a factor of 3-4, signaling strong contributions from low-lying resonances. In this work, we provide a theoretical analysis of $\Lambda_c^+ \to n \bar{K}^0\pi^+$ within the coupled-channel chiral unitary approach, where the $N(1535)$ and $\Lambda(1670)$ can be dynamically generated. Our calculations show a narrow peak from $N(1535)$ in the $\pi^+ n$ invariant mass spectrum and a distinct dip from $\Lambda(1670)$ in the $\bar{K}^0 n$ spectrum. The dip structure is qualitatively consistent with the $\Lambda(1670)$ manifestation in $\bar{K}N \to \bar{K}N$ scattering, supporting its molecular interpretation. This study not only connects the experimental observations but also highlights $\Lambda_c^+ \to n \bar{K}^0\pi^+$ as a crucial process to disentangle the nature of $N(1535)$ and $\Lambda(1670)$. Future precise measurements of this decay channel by the BESIII, Belle II, LHCb, and the proposed Super Tau-Charm Factory are strongly encouraged.

\end{abstract}

\maketitle

\section{Introduction} \label{sec:Introduction}
Hadronic decays of charmed baryons serve as powerful laboratories for probing the structure of low-lying baryon resonances and the dynamics of final-state interactions~\cite{BESIII:2025rda,Belle:2025voy,Wang:2024jyk,Feng:2020jvp,Zeng:2020och,Wang:2020pem}. Among these, the decay $\Lambda_c^+ \to n \bar{K}^0\pi^+$ (or its neutral-kaon counterpart $\Lambda_c^+ \to n K_S^0 \pi^+$) has recently drawn significant attention due to intriguing experimental findings and its unique sensitivity to specific resonance contributions~\cite{BESIII:2016yrc,BESIII:2023pia}.

The process $\Lambda_c^+ \to n K_S^0 \pi^+$ has been precisely measured by the Beijing Spectrometer III (BESIII) Collaboration, yielding a branching fraction of $(1.86 \pm 0.08 \pm 0.04) \times 10^{-2}$~\cite{BESIII:2023pia}. This value is notably 3 to 4 times larger than theoretical predictions based on SU(3) flavor symmetry~\cite{Geng:2018upx,Cen:2019ims}, strongly suggesting the presence of significant resonant contributions beyond simple spectator diagrams. Since contributions from excited kaons are Double Cabibbo-suppressed in this decay, the dominant resonant enhancement likely stems from low-lying baryon states. Two prime candidates are the $N(1535)$ ($J^P=1/2^-$) and the $\Lambda(1670)$ ($J^P=1/2^-$), both of which couple strongly to channels involving strangeness.

However, the situation is complicated by an apparent discrepancy concerning the isospin composition of the $\bar{K}N$ system in $\Lambda_c^+$ decays. In the analysis of $\Lambda_c^+ \to p K^-\pi^+$, both the LHCb~\cite{LHCb:2022sck} and Belle~\cite{Belle:2022cbs} Collaborations found that the amplitude describing the $p K^-$ ($I=0$) pair is dominant, supported by the studies of Refs.~\cite{Zhang:2024jby,Duan:2024okk}. In contrast, BESIII's analysis of $\Lambda_c^+ \to n K_S^0 \pi^+$ suggests that the $n\bar{K}^0$ system receives contributions from both isospin $0$ and isospin $1$ components, with comparable importance~\cite{BESIII:2023pia,Lu:2016ogy}. This inconsistency calls for a deeper theoretical investigation that connects these decay modes and clarifies the role of intermediate resonances, which can significantly alter the effective isospin projection observed in different final states.

The $N(1535)$ resonance, the lowest-lying $J^P=1/2^-$ nucleon excitation, has long been a subject of debate regarding its internal structure~\cite{Liu:2005pm,Geng:2008cv}. In the traditional quark model, it is described as a three-quark orbital excitation. However, its unexpectedly large couplings to $\eta N$ and $K\Lambda$ channels has motivated alternative interpretations, such as a significant pentaquark component $[ud][us]\bar{s}$~\cite{Hannelius:2000gu,Helminen:2000jb,Zhang:2004xt,Zou:2007mk} or a dynamically generated state from meson-baryon interactions within the chiral unitary approach~\cite{Oset:1997it,Jido:2003cb,Kaiser:1996js,Nieves:2001wt,Inoue:2001ip,Doring:2008sv,Wang:2015pcn,Lu:2016roh, Pavao:2018wdf,Lyu:2023aqn,Xie:2017erh,Li:2024rqb,Li:2025gvo,Song:2025eko,Li:2026lbo}. Within the Hamiltonian effective field theory, the $N(1535)$ resonance can be interpreted as a three‑quark core generated from the meson–baryon scattering~\cite{Liu:2015ktc,Guo:2022hud,Abell:2023nex}. Recent studies using correlation functions and scattering lengths also explore its possible molecular nature~\cite{Molina:2023jov,Li:2023pjx}. Thus, $\Lambda_c^+ \to n \bar{K}^0\pi^+$, where the $\pi^+ n$ spectrum provides a relatively clean environment for $N(1535)$ (free from double Cabibbo-suppressed kaon excitations), is an ideal process to study this contested state.

The $\Lambda(1670)$ resonance presents another fascinating puzzle due to its process-dependent line shapes. While it appears as a peak in the $\eta\Lambda$ spectrum of $\Lambda_c^+ \to \Lambda \pi^+ \eta$~\cite{Belle:2020xku} and as a cusp near the $\eta\Lambda$ threshold in the $K^- p$ spectrum of $\Lambda_c^+ \to p K^-\pi^+$~\cite{LHCb:2022sck,Belle:2022cbs}, it famously manifests as a dip in the $\bar{K}N$ invariant mass distribution of the $\bar{K}N \to \bar{K}N$ scattering process~\cite{Gopal:1976gs,Oset:2001cn}. 
On the other hand, the Crystal Ball Collaboration has measured the differential and total cross-section for the process $K^-p \to \eta\Lambda$, suggesting that the $\Lambda(1670)$ is a three-quark state~\cite{CrystalBall:2001uhc}. The study of the $K^-p \to \pi^0\Lambda^0$ scattering process within the chiral quark model, supports the $\Lambda(1670)$ as a traditional three-quark state~\cite{Zhong:2008km}. In Ref.~\cite{Liu:2023xvy}, the authors investigated the internal structure of $\Lambda(1670)$ within the Hamiltonian effective field theory by combining lattice QCD simulations and experimental data, and prefer to describe $\Lambda(1670)$ as a bare three-quark basis state that mixes with the $\pi\Sigma$, $\bar{K}N$, $\eta\Lambda$, and $K\Xi$ meson-baryon channels. 
Thus, this transition from a peak to a cusp/dip across different production mechanisms highlights the sensitivity of its observable signature to interference with non-resonant backgrounds and the specific coupled channels involved.
Theoretical interpretations of $\Lambda(1670)$ also vary, with descriptions ranging from a conventional three-quark state~\cite{CrystalBall:2001uhc,Zhong:2008km,Liu:2023xvy} to a dynamically generated molecule from coupled-channel interactions like $\bar{K}N$, $\eta\Lambda$, and $\pi\Sigma$~\cite{Xie:2016evi,Wang:2022nac,Miyahara:2015cja,Zhang:2024jby,Duan:2024okk,Duan:2024czu,Lyu:2024qgc,Wang:2022xqc}. The $\bar{K}^0 n$ spectrum in $\Lambda_c^+ \to n \bar{K}^0\pi^+$ offers a new terrain to map the $\Lambda(1670)$ line shape and test these interpretations.

In this work, we perform a theoretical analysis of the decay $\Lambda_c^+ \to n \bar{K}^0\pi^+$ within the framework of the chiral unitary approach, where the $N(1535)$ and $\Lambda(1670)$ resonances are dynamically generated from $S$-wave pseudoscalar meson-octet baryon final-state interactions (FSI). Our primary goals are: (i) to provide theoretical predictions for the $\pi^+n$ and $\bar{K}^0n$ invariant mass distributions, highlighting the characteristic structures from $N(1535)$ and $\Lambda(1670)$; (ii) to discuss how these resonant contributions can be helpful to reconcile the observed experimental discrepancies in $\bar{K}N$ isospin composition; and (iii) to emphasize the role of this decay channel as a unique probe for elucidating the nature of these two enigmatic low-lying resonances. 

The paper is organized as follows. In Sec.~\ref{sec:Formalism}, we detail the theoretical framework, including the weak production mechanisms, the hadronization process, and the FSI leading to the dynamical generation of $N(1535)$ and $\Lambda(1670)$. In Sec.~\ref{sec:Results}, we present and discuss our numerical results for the invariant mass distributions, the Dalitz plot, and the dependence on model parameters. A summary and outlook are given in Sec.~\ref{sec:Conclusions}.

\section{Formalism} \label{sec:Formalism}

In this section, we first present the theoretical formalism for the process $\Lambda_c^+ \to n \bar{K}^0\pi^+$. The decay process proceeds through three sequential steps: weak decay, hadronization, and final-state interaction. Then, we show the amplitude of the dynamically generated $N(1535)$ in Sec.~\ref{sec2a} and that of the dynamically generated  $\Lambda(1670)$ in Sec.~\ref{sec2b}. Finally, we give the invariant mass distributions of the process $\Lambda_c^+ \to n \bar{K}^0\pi^+$ in Sec.~\ref{sec2c}.

\begin{figure}[tbhp]
	\centering
	\includegraphics[scale=0.6]{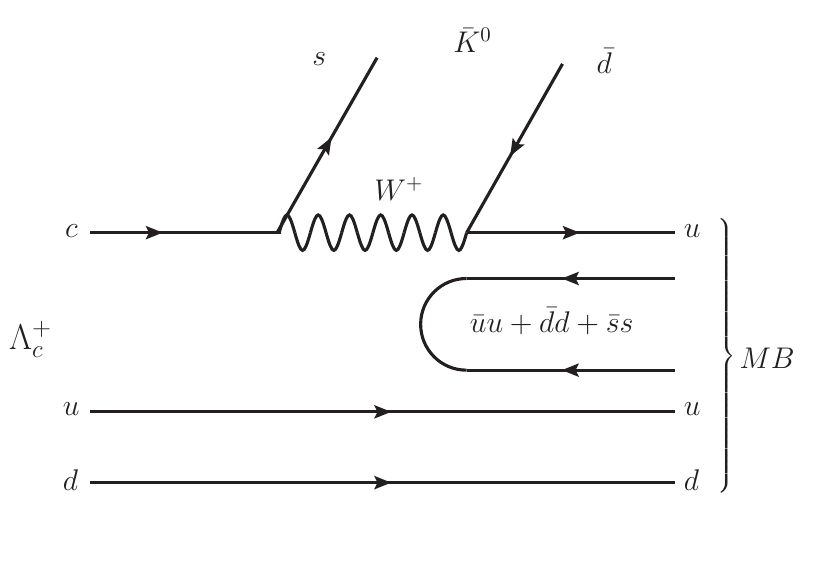}
	\caption{Quark level diagram for the process $\Lambda_c^+ \to  \bar{K}^0MB$ via the $W^+$ internal emission.}
	\label{fig:level_pi+n}
\end{figure}

The quark-level diagram for the internal emission mechanism of $\Lambda_c^+ \to n \bar{K}^0\pi^+$ is shown in Fig.~\ref{fig:level_pi+n}, as done in Refs.~\cite{Zhang:2026igc,Li:2025exm}. The process begins with the weak decay $c \to s W^+$, followed by $W^+ \to u\bar{d}$. Subsequently, the $s$ quark and the $\bar{d}$ quark from the ${W^{+}}$ boson will hadronize into $\bar{K}^0$, while the $u$ quark, the spectator $ud$ diquark from the initial $\Lambda_c^+$, and a $\bar{u}u+\bar{d}d+\bar{s}s$ pair created from the vacuum,  hadronize into a meson-baryon pair. The flavor structure for the weak decay and hadronization are given below,
\begin{align}
	\Lambda_c^+&= \frac{1}{\sqrt{2}}c(ud-du) \nonumber\\
	&\Rightarrow V_{cs}V_{ud} \frac{1}{\sqrt{2}}s\bar{d}u\left(\bar{u}u+\bar{d}d+\bar{s}s\right)\left(ud-du\right) \nonumber\\
	&=V_{cs}V_{ud}\frac{1}{\sqrt{2}}\bar{K}^0u\left(\bar{u}u+\bar{d}d+\bar{s}s\right)\left(ud-du\right), 
\end{align}
where $V_{cs}$ and $V_{ud}$ denote the Cabibbo-Kobayashi-Maskawa (CKM) matrix elements for the quark transitions $c \to s$ and $u \to d$, respectively. Within the SU(3) flavor symmetry, the quark and hadron degrees of freedom could be correlated by the ground pseudoscalar meson matrix $M$ and the ground baryon octet matrix $B$, and the matrices $M$ and $B$ are expressed as~\cite{Miyahara:2016yyh,Lyu:2023aqn,Pavao:2017cpt},
\begin{align} 
	M &= 
	\renewcommand{\arraystretch}{1.2}  
	\begin{pmatrix}
		u\bar{u} & \hspace{0.3em} u\bar{d} & \hspace{0.3em} u\bar{s} \\  
		d\bar{u} & \hspace{0.3em} d\bar{d} & \hspace{0.3em} d\bar{s} \\
		s\bar{u} & \hspace{0.3em} s\bar{d} & \hspace{0.3em} s\bar{s}
	\end{pmatrix} \nonumber\\
	&= 
	\renewcommand{\arraystretch}{1.2}
	\begin{pmatrix}
		\frac{\eta}{\sqrt{3}} + \frac{\pi^0}{\sqrt{2}} + \frac{\eta'}{\sqrt{6}} & \hspace{0.3em} \pi^+ & \hspace{0.3em} K^+ \\
		\pi^- & \hspace{0.3em} \frac{\eta}{\sqrt{3}} - \frac{\pi^0}{\sqrt{2}} + \frac{\eta'}{\sqrt{6}} & \hspace{0.3em} K^0 \\
		K^- & \hspace{0.3em} \bar{K}^0 & \hspace{0.3em} -\frac{\eta}{\sqrt{3}} + \frac{\sqrt{6}\eta'}{3}
	\end{pmatrix}, \label{M_matrix} \\  
	B &= \frac{1}{\sqrt{2}}
	\renewcommand{\arraystretch}{1.2}
	\begin{pmatrix}
		u(ds - sd) & \hspace{0.3em} u(su - us) & \hspace{0.3em} u(ud - du) \\
		d(ds - sd) & \hspace{0.3em} d(su - us) & \hspace{0.3em} d(ud - du) \\
		s(ds - sd) & \hspace{0.3em} s(su - us) & \hspace{0.3em} s(ud - du)
	\end{pmatrix} \nonumber\\
	&= 
	\renewcommand{\arraystretch}{1.2}
	\begin{pmatrix}
		\frac{\Sigma^0}{\sqrt{2}} + \frac{\Lambda}{\sqrt{6}} & \hspace{0.3em} \Sigma^+ & \hspace{0.3em} p \\
		\Sigma^- & \hspace{0.3em} -\frac{\Sigma^0}{\sqrt{2}} + \frac{\Lambda}{\sqrt{6}} & \hspace{0.3em} n \\
		\Xi^- & \hspace{0.3em} \Xi^0 & \hspace{0.3em} -\frac{2\Lambda}{\sqrt{6}}
	\end{pmatrix}, \label{B_matrix}
\end{align}
where we adopt the $\eta$-$\eta^{\prime}$ standard mixing~\cite{Bramon:1992kr}. Given that the $\eta^{\prime}$ has a large mass and is typically omitted in chiral Lagrangians~\cite{Oset:2016lyh}, we neglect the meson-baryon pair involving the $\eta^{\prime}$. Then, the available final $|MB\rangle$ pairs are written as,
\begin{align}
	|MB\rangle &= \frac{1}{\sqrt{2}} |u(\bar{u}u + \bar{d}d + \bar{s}s)(ud - du)\rangle  \nonumber \\
	&= \sum_{i=1}^{3} |M_{1i}B_{i3}\rangle \nonumber \\
	&= \frac{1}{\sqrt{2}} |\pi^0 p\rangle + \frac{1}{\sqrt{3}} |\eta p\rangle + |\pi^+ n\rangle - \sqrt{\frac{2}{3}} |K^+ \Lambda\rangle,
\end{align}
where $i= 1, 2, 3$ corresponds to the $u$, $d$, and $s$ quarks, respectively. Thus, the total amplitude for the internal emission mechanism contributing to $\Lambda_c^+\to \bar{K}^0X$, where $X$ later undergoes FSI, is,
\begin{align}\label{lambda-pin}
	\mathcal{M}^{\rm in} = V_p V_{cs}V_{ud} \bar{K}^{0}\left(\frac{1}{\sqrt{2}}\pi^{0}p+\frac{1}{\sqrt{3}}\eta p+\pi^+ n-\sqrt{\frac{2}{3}}K^{+}\Lambda\right),
\end{align}
where $V_p$ is a constant factor encoding the weak transition strength.

\begin{figure}[tbhp]
	\centering
	\includegraphics[scale=0.6]{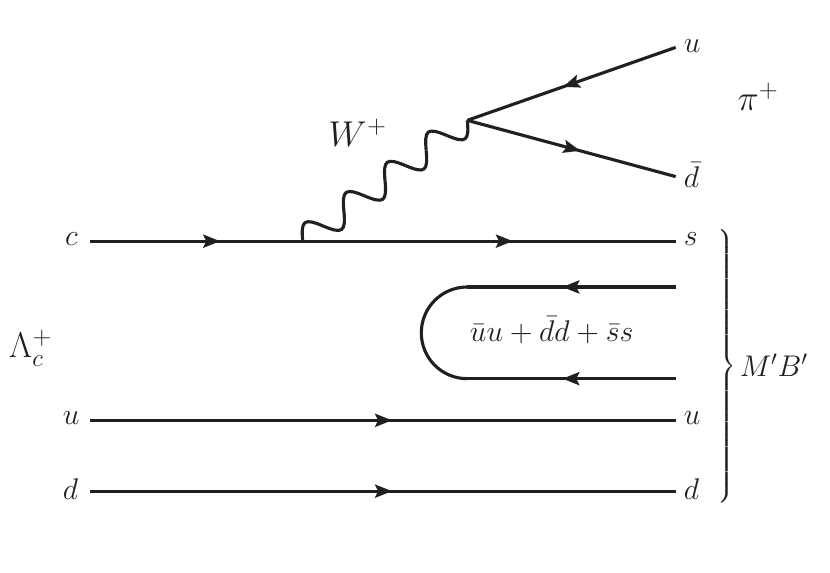}
	\caption{Quark level diagram for the process $\Lambda_c^+ \to \pi^+ M^{\prime}B^{\prime}$ via the $W^+$ external emission.}
	\label{fig:level_kn}
\end{figure}
Subsequently, we present the quark-level diagram for the external emission mechanism of the decay $\Lambda_c^+ \to n \bar{K}^0\pi^+$ in Fig.~\ref{fig:level_kn}. In the first step, the charm quark of initial $\Lambda_c^+$ decays into a strange quark and a $u\bar{d}$ pair by the weak interaction, then the $u\bar{d}$ quark pair hadronizes into a $\pi^+$. Meanwhile, the $sud$ cluster hadronizes with the antiquark-quark pair $\bar{u}u+\bar{d}d+\bar{s}s$ created from the vacuum, and we could obtain,
\begin{align}
	\Lambda_c^+&= V_{cs}V_{ud} \frac{1}{\sqrt{2}}c(ud-du) \nonumber\\
	&\Rightarrow V_{cs}V_{ud} \frac{1}{\sqrt{2}}u\bar{d}s\left(\bar{u}u+\bar{d}d+\bar{s}s\right)\left(ud-du\right) \nonumber\\
	&=V_{cs}V_{ud} \frac{1}{\sqrt{2}}\pi^+s\left(\bar{u}u+\bar{d}d+\bar{s}s\right)\left(ud-du\right) .
\end{align}
The resulting meson-baryon state $|M^{\prime}B^{\prime}\rangle$ in Fig.~\ref{fig:level_kn} can be written as,
\begin{align}
	|M^{\prime}B^{\prime}\rangle &= \frac{1}{\sqrt{2}} |s(\bar{u}u + \bar{d}d + \bar{s}s)(ud - du)\rangle  \nonumber \\
	&= \sum_{i=1}^{3} |M_{3i}^{\prime}B_{i3}^{\prime}\rangle \nonumber \\
	&= |K^- p\rangle + |\bar{K}^0 n\rangle + \frac{\sqrt{2}}{3} |\eta \Lambda\rangle.
\end{align}
The external emission amplitude is therefore,
\begin{align}\label{lambda-kn}
	\mathcal{M}^{\rm ex} = C V_p V_{cs}V_{ud} \pi^{+}\left(K^- p+\bar{K}^0n+\frac{\sqrt{2}}{3}\eta \Lambda\right),
\end{align}
where the color factor $C$ accounts for the relative weight of the external vs.\ internal emission diagrams. Both mechanisms are Cabibbo-favored and of the same order.

\subsection{Dynamical Generation of $N(1535)$}\label{sec2a}

\begin{figure}[tbhp]
	\centering
	\includegraphics[scale=0.43]{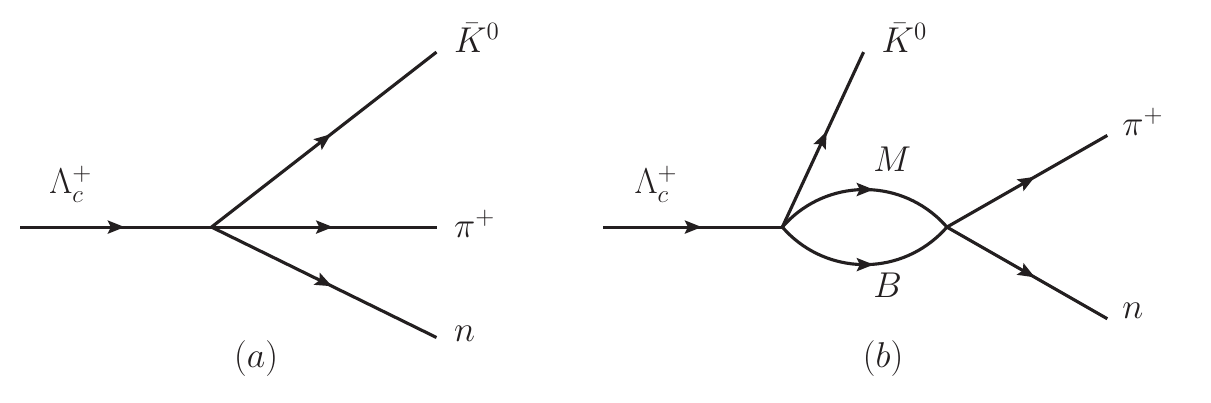}
	\caption{The diagrams for the Tree-level (a) and the $MB$ rescattering (b) of the $\Lambda_c^+ \to  n \bar{K}^0\pi^+$ decay.}
	\label{fig:pi+n_quan}
\end{figure}
As depicted in Fig.~\ref{fig:pi+n_quan}(a), the final state $n \bar{K}^0\pi^+$ could be generated directly. Furthermore, this final state may also arise from the rescattering of intermediate meson-baryon pair in $S$-wave, as illustrated in Fig.~\ref{fig:pi+n_quan}(b), which could dynamically generate the $N(1535)$. 
As we discussed in the introduction, the $N(1535)$ resonance is dynamically generated from the $S$-wave interaction of pseudoscalar mesons with octet baryons in the strangeness $S=0$, isospin $I=1/2$ sector. We consider the coupled channels: $\pi^0 p$, $\pi^+ n$, $\eta p$, $K^+ \Lambda$, $K^0 \Sigma^+$, and $K^+ \Sigma^0$.
Based on the final-state components given in Eqs.~\eqref{lambda-pin} and \eqref{lambda-kn}, the decay amplitude corresponding to the dynamically generated $N(1535)$ resonance could be expressed as~\footnote{The CKM matrix elements $V_{cs}V_{ud}$ have been absorbed in the constant $V_p$ in the following.},
\begin{align} \label{Eq:M-N1535}
		\mathcal{T}^{N(1535)} =&\; V_p \bigg( h_{\pi^+ n} 
		+ h_{\pi^0 p}\tilde{G}_{\pi^0 p}(M_{\pi^+ n}) t_{\pi^0 p \to \pi^+ n}(M_{\pi^+ n}) \notag \\
		&+ h_{\eta p} \tilde{G}_{\eta p}(M_{\pi^+ n}) t_{\eta p \to \pi^+ n}(M_{\pi^+ n}) \notag \\
		&+ h_{\pi^+ n} \tilde{G}_{\pi^+ n}(M_{\pi^+ n}) t_{\pi^+ n \to \pi^+ n}(M_{\pi^+ n}) \notag \\
		&+ h_{K^+ \Lambda} \tilde{G}_{K^+ \Lambda}(M_{\pi^+ n}) t_{K^+ \Lambda \to \pi^+ n}(M_{\pi^+ n}) \bigg)  \notag \\
		&+ C V_p  \bigg( h_{\bar{K}^0 n} \tilde{G}_{\pi^+ n}(M_{\pi^+ n}) t_{\pi^+ n \to \pi^+ n}(M_{\pi^+ n}). \bigg)  
\end{align}
where the coefficients $h_{MB}$ represent the weights of the respective coupling channels in the hadroniation process, and we can obtain from Eqs.~\eqref{lambda-pin} and \eqref{lambda-kn},
\begin{equation}
	h_{\pi^0 p} =\frac{1}{\sqrt{2}}, \; h_{\eta p}=\frac{1}{\sqrt{3}}, \; h_{\pi^+ n}=1, \;  h_{K^+ \Lambda}=-\sqrt{\frac{2}{3}},\; h_{\bar{K}^0 n}=1.
\end{equation}
The coupled channels transition amplitude of $t_{MB \to \pi^+ n}$ in Eq.~(\ref{Eq:M-N1535}) can be obtained by solving the Bethe-Salpeter equation
\begin{equation}\label{Eq:BS}
	T=[1-VG]^{-1}V,
\end{equation}
where $G$ is the meson-baryon loop function, and the interaction kernel $V$ is taken from the leading order of the chiral Lagrangian, while we take it from Ref.~\cite{Wang:2015pcn}
\begin{align} \label{Eq:vij-pin}
	V_{ij} &= -C_{ij} \frac{1}{4f_i f_j} \bigl(2\sqrt{s} - M_i - M_j\bigr) \notag \\
	&\quad \times \left( \frac{M_i + E_i}{2M_i} \right)^{1/2} \left( \frac{M_j + E_j}{2M_j} \right)^{1/2},
\end{align}   
where $M_{i(j)}$ and $E_{i(j)}$ are the mass and energy of the baryon in the $i(j)$ channel, with $E_{i}=({s+M_{i}^2-m_{i}^2})/{2\sqrt{s}}$. The coefficients $C_{ij}$ are taken from  Table~\ref{tab:Cij-pin}, with the pseudoscalar decay constants $f_{i(j)}$ as follows,
\begin{equation}
	f_\pi=93~\mathrm{MeV},\quad f_K=1.22f_\pi,\quad f_\eta=1.3f_\pi.
\end{equation}

The meson-baryon system loop function $G$ in Eq.~(\ref{Eq:BS}) is given by~\cite{Inoue:2001ip},
\begin{equation}\label{loop function}
G_{i}=i\int\frac{d^4q}{(2\pi)^4}\frac{2M_i}{(P-q)^2-M_i^2+i\epsilon}\frac{1}{q^2-m_i^2+i\epsilon},
\end{equation}
where $M_{i}$($m_{i}$) is the baryon(meson) mass for the $i$-th coupled channel. $P$ is the four-momentum of the meson-baryon system, and $q$ is the meson four-momentum in the center-of-mass (c.m.) frame. The loop function $G_{i}$ in Eq.~(\ref{loop function}) is divergent, and we typically adopt either the three-momentum cutoff method or the dimensional regularization method for its renormalization. In this work, we employ dimensional regularization method, and the corresponding meson-baryon loop function $G_{MB}$ can be written as~\cite{Wang:2015pcn,Li:2025exm},  
\begin{align}  \label{loop function-MB}
	G(MB) &= \frac{2M_B}{16\pi^2} \Bigg\{ a_{MB}(\mu) + \ln\frac{M_B^2}{\mu^2} + \frac{m_M^2 - M_B^2 + s}{2s} \ln\frac{m_M^2}{M_B^2} \notag \\
	&\quad + \frac{q_{MB}}{\sqrt{s}} \bigg[ \ln\big(s - (M_B^2 - m_M^2) + 2q_{MB}\sqrt{s}\big) \notag \\
	&\quad + \ln\big(s + (M_B^2 - m_M^2) + 2q_{MB}\sqrt{s}\big) \notag \\
	&\quad - \ln\big(-s + (M_B^2 - m_M^2) + 2q_{MB}\sqrt{s}\big) \notag \\
	&\quad - \ln\big(-s - (M_B^2 - m_M^2) + 2q_{MB}\sqrt{s}\big) \bigg] \Bigg\},
\end{align}
where $\sqrt{s}$ is the invariant mass of the $MB$ pair, $m_M$ and $M_B$ are the meson and baryon masses for the coupled channels, respectively. The $q_{MB}$ is the meson momentum in the meson-baryon c.m. frame, with $q_{MB}={\lambda^{1/2}(s,m_{M}^2,M_{B}^2)}/{2\sqrt{s}}$. The regularization scale $\mu$ and the subtraction constant $a_{MB}(\mu)$ are determined by fitting to the experimental data. The values of $\mu$ and $a_i(\mu)$ presented in Ref.~\cite{Inoue:2001ip} are exactly a set of parameters obtained by fitting to the experimental data of elastic $\pi N$ $T$ matrix and $\pi N\to \pi \pi N$ cross section. Following Ref.~\cite{Inoue:2001ip}, we take $\mu =1200$~MeV for each channel and the values for the subtraction constants are as follows:
\begin{align}  \label{subtraction constants}
	&a_{K^0\Sigma^+}(\mu) = -2.8, \quad a_{K^+\Sigma^0}(\mu) = -2.8, \quad a_{K^+\Lambda}(\mu) = 1.6, \notag \\
	&a_{\pi^+ n}(\mu) = 2.0, \quad\quad\; a_{\pi^0p}(\mu) = 2.0, \quad\quad\; a_{\eta p}(\mu) = 0.2.
\end{align}

It should be noted that when calculating the loop function in Eq.~(\ref{loop function}), matching the result obtained using the three-momentum cutoff method with that from the dimensional regularization method in Eq.~(\ref{loop function-MB}) will yield negative values for the subtraction constants $a_i(\mu)$. However, the subtraction constants for the $K^+\Lambda$, $\pi^+ n$, $\pi^0 p$, and $\eta p$ channels in Eq.~(\ref{subtraction constants}) are positive. Ref.~\cite{Inoue:2001ip} introduced these odd values in order to account for the missing channels in the scattering amplitude~\cite{Hyodo:2008xr}. Since the $MB$ rescattering in Fig.~\ref{fig:pi+n_quan}(b) involves only four channels, we are not justified in using the loop function $G$ in the scattering amplitude to account for channels that do not contribute to this process. As done in Refs.~\cite{Wang:2015pcn,Lu:2016roh}, the meson-baryon loop function $\tilde{G}$ in Eq.~(\ref{Eq:M-N1535}) employs the cutoff method, rather than the dimensional regularization, and we take a cutoff momentum  of $|\tilde{q}_{\text{max}}| = 1300$~MeV in our calculation. The form of $\tilde{G}$ is given by, 
\begin{align}
	\tilde{G}_{i}(M_{\pi^{+}n}) & =\int\frac{d^3q}{(2\pi)^3}\frac{M_i}{2\omega_i(q)E_i(q)} \notag \\
	& \times\frac{1}{M_{\pi^+n}-\omega_i(q)-E_i(q)+\mathrm{i}\epsilon},\label{eq:gfunction_cutoff}
\end{align}
where $M_i$, $E_i$, and $\omega_i$ are the baryon mass, baryon energy and meson energy of the $i$-th channel.

\begin{table}[htbp]
	\begin{center}    
		\caption{Coefficients $C_{ij}$ for the $S$-wave meson-baryon $(MB)$ scattering~\cite{Doring:2005bx}.}     \label{tab:Cij-pin}
		\renewcommand{\arraystretch}{1.5}
		\setlength{\tabcolsep}{6.2pt}
		\begin{tabular}{ccccccc}
			\hline
			\hline
			& $K^0\Sigma^+$ & $K^+\Sigma^0$ & $K^+\Lambda$ & $\pi^+n$ & $\pi^0p$ & $\eta p$ \\
			\hline
			$K^0\Sigma^+$ & $1$ & $\sqrt{2}$ & $0$ & $0$ & $1/\sqrt{2}$ & $-\sqrt{{3}/{2}}$ \\
			$K^+\Sigma^0$ &  & $0$ & $0$ & ${1}/{\sqrt{2}}$ & $-{1}/{2}$ & $-{\sqrt{3}}/{2}$ \\
			$K^+\Lambda$ &  &  & $0$ & $-\sqrt{{3}/{2}}$ & $-{\sqrt{3}}/{2}$ & $-{3}/{2}$ \\
			$\pi^+n$ &  &  &  & $1$ & $\sqrt{2}$ & $0$ \\
			$\pi^0p$ &  &  &  &  & $0$ & $0$ \\
			$\eta p$ &  &  &  &  &  & $0$ \\
			\hline
			\hline
		\end{tabular}
		\renewcommand{\arraystretch}{1.5}
		\setlength{\tabcolsep}{6.2pt}
	\end{center}
\end{table}

\subsection{Dynamical Generation of $\Lambda(1670)$ }\label{sec2b}
\begin{figure}[tbhp]
	\centering
	\includegraphics[scale=0.43]{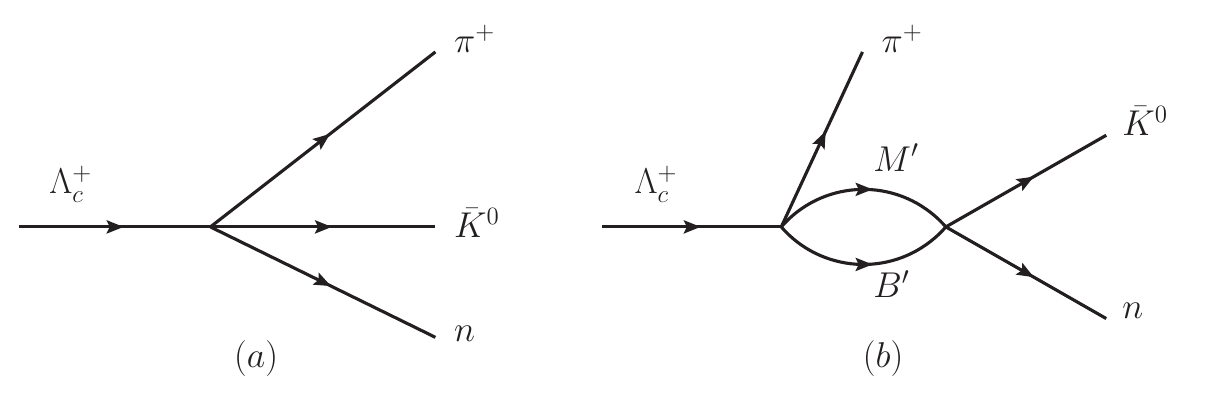}
	\caption{The diagrams for the Tree-level (a) and the $M^{\prime}B^{\prime}$ rescattering (b) of the $\Lambda_c^+ \to  n \bar{K}^0\pi^+$ decay.}
	\label{fig:kn_quan}
\end{figure}

In this subsection, we investigate the $\bar{K}^0 n$ final-state interaction which dynamically generates the $\Lambda(1670)$ resonance as shown in Fig.~\ref{fig:kn_quan}. The mechanism includes both tree-level and rescattering contributions to the decay.  Based on Eqs.~\eqref{lambda-pin} and \eqref{lambda-kn}, the decay amplitude for the $\Lambda(1670)$ contribution to $\Lambda_c^+\to n \bar{K}^0 \pi^+$ could be expressed as 
\begin{align}   \label{Eq:M-1670}
	\mathcal{T}^{\Lambda(1670)} =& \; C V_p \bigg(  h_{\bar{K}^0 n} + h_{K^- p} G_{K^- p}(M_{\bar{K}^0 n}) t_{K^- p \to \bar{K}^0 n}(M_{\bar{K}^0 n}) \notag \\
	&+ h_{\bar{K}^0 n} G_{\bar{K}^0 n}(M_{\bar{K}^0 n}) t_{\bar{K}^0 n \to \bar{K}^0 n}(M_{\bar{K}^0 n}) \notag \\
	&+ h_{\eta \Lambda} G_{\eta \Lambda}(M_{\bar{K}^0 n}) t_{\eta \Lambda \to \bar{K}^0 n}(M_{\bar{K}^0 n}) \bigg) \notag \\
	&+ V_p \bigg(h_{\pi^+ n} G_{\bar{K}^0 n}(M_{\bar{K}^0 n}) t_{\bar{K}^0 n \to \bar{K}^0 n}(M_{\bar{K}^0 n}) \bigg)
\end{align}
with $h_{K^- p}=1$ and $h_{\eta \Lambda}={\sqrt{2}}/{3}$. The meson-baryon loop function $G$ follows the same form as Eq.~(\ref{loop function-MB}), while the meson-baryon transition amplitude $t_{M^{\prime}B^{\prime} \to \bar{K}^0 n}$ is likewise obtained through the Bethe–Salpeter equation as Eq.~(\ref{Eq:BS}). 
The $\Lambda(1670)$ resonance is dynamically generated in the strangeness $S=-1$, isospin $I=0$ sector. We consider ten coupled channels: $K^- p$, $\bar{K}^0 n$, $\pi^0 \Lambda$, $\pi^0 \Sigma^0$, $\eta \Lambda$, $\eta \Sigma^0$, $\pi^+ \Sigma^-$, $\pi^- \Sigma^+$, $K^+ \Xi^-$, and $K^0 \Xi^0$. Here, we take $\mu = 630$~MeV, and the subtraction constants for each coupled channel are $a_{\bar{K}N}=-1.84$, $a_{\pi\Sigma}=-2.00$, $a_{\pi\Lambda}=-1.83$, $a_{\eta\Lambda}=-2.25$, $a_{\eta\Sigma}=-2.38$, and $a_{K\Xi}=-2.52$, taken from Ref.~\cite{Oset:2001cn}.

The transition potentials $V_{ij}^{\prime}$ between the ten channels are given by~\cite{Oset:2001cn},
\begin{align}
	V_{ij}^{\prime} &= -C_{ij}^{\prime} \frac{1}{4 f^2} \bigl(2\sqrt{s} - M_i - M_j\bigr) \notag \\
	&\quad \times \left( \frac{M_i + E_i}{2M_i} \right)^{1/2} \left( \frac{M_j + E_j}{2M_j} \right)^{1/2},
\end{align}
where $f$=$1.15 f_{\pi}$, and the coefficients $C_{ij}^{\prime}$ are given in Table~I of Ref.~\cite{Oset:1997it}. 

It should be pointed out that the pole position of $\Lambda(1670)$ is highly sensitive to the parameter $a_{K\Xi}$, while it is only moderately sensitive to the subtraction constants $a_{\bar{K}N}$, $a_{\pi\Sigma}$, and $a_{\eta\Lambda}$. In Refs.~\cite{Wang:2022nac,Lyu:2024qgc,Zhang:2024jby}, the authors determine $a_{K\Xi}$ by fitting to the experimental data. Since no experimental data are provided for the decay $\Lambda_c^+ \to n \bar{K}^0\pi^+$, we adopt $a_{K\Xi}=-2.776$ from Ref.~\cite{Zhang:2024jby}. Future measurements of $\Lambda_c^+ \to n\bar{K}^0 \pi^+ $ could help to constrain this parameter.

\subsection{Invariant Mass Distributions}\label{sec2c}
Based on the theoretical formalisms mentioned above, one can write down the double differential width of the process $\Lambda_c^+ \to n \bar{K}^0\pi^+$ as
\begin{eqnarray}
    \frac{d^{2}\Gamma}{dM_{\pi^+ n} dM_{\bar{K}^0 n}}&=\dfrac{1}{(2\pi)^{3}}\dfrac{M_{n}}{2M_{\Lambda_c^+}^2}|\mathcal{T}|^{2} M_{\pi^+ n} M_{\bar{K}^0 n}, \label {eq:dgammadm12dm23} 
\end{eqnarray}
 where $M_{\pi^+ n}$ and $M_{\bar{K}^0 n}$ are the invariant masses of the indicated pairs, respectively. $\mathcal{T}$ is the total decay amplitude for the decay $\Lambda_c^+ \to n \bar{K}^0\pi^+$, 
\begin{eqnarray}\label{Eq:M-total}
\mathcal{T}=\mathcal{T}^{N(1535)}+\mathcal{T}^{\Lambda(1670)}\times e^{i\phi},
\end{eqnarray}
including a possible relative phase angle $\phi$.

One could easily obtain the single differential decay widths $d\Gamma/dM_{\pi^+ n}$ and $d\Gamma/dM_{\bar{K}^0 n}$ by integrating over the other invariant mass within its kinematically allowed range. For a fixed $M_{12}$, the corresponding range for the  $M_{23}$ is~\cite{ParticleDataGroup:2024cfk},
\begin{align}
	&\left(M_{23}\right)_{\min}=\sqrt{\left(E_2^*+E_3^*\right)^2-\left(\sqrt{E_2^{* 2}-m_2^2}+\sqrt{E_3^{* 2}-m_3^2}\right)^2}, \nonumber\\
	&\left(M_{23}\right)_{\max}=\sqrt{\left(E_2^*+E_3^*\right)^2-\left(\sqrt{E_2^{* 2}-m_2^2}-\sqrt{E_3^{* 2}-m_3^2}\right)^2}, \label{eq:limit}
\end{align}
where $E_2^*$ and $E_3^*$ are the energies of particles 2 and 3 in the rest frame of $M_{12}$, respectively,
\begin{align}
	E_{2}^{*}&=\frac{M_{12}^{2}-m_{1}^{2}+m_{2}^{2}}{2M_{12}}, \nonumber\\
	E_{3}^{*}&=\frac{M_{\Lambda_c^{+}}^{2}-M_{12}^{2}-m_{3}^{2}}{2M_{12}},
\end{align}
where $m_1$, $m_2$, and $m_3$ represent the masses of particles 1, 2, and 3,  respectively. The masses and widths of the relevant particles are taken from the Review of Particle Physics (RPP)~\cite{ParticleDataGroup:2024cfk}.
Since absolute experimental distributions are not yet available, we set the overall normalization $V_p = 1$ and the color factor $C = 3$ as a reference. Our predictions focus on the line shapes, which are independent of this overall normalization.


\section{Results and Discussion} \label{sec:Results}

\subsection{Transition Amplitudes and Resonance Signals}

\begin{figure}[tbhp]
	\centering
	\includegraphics[scale=0.7]{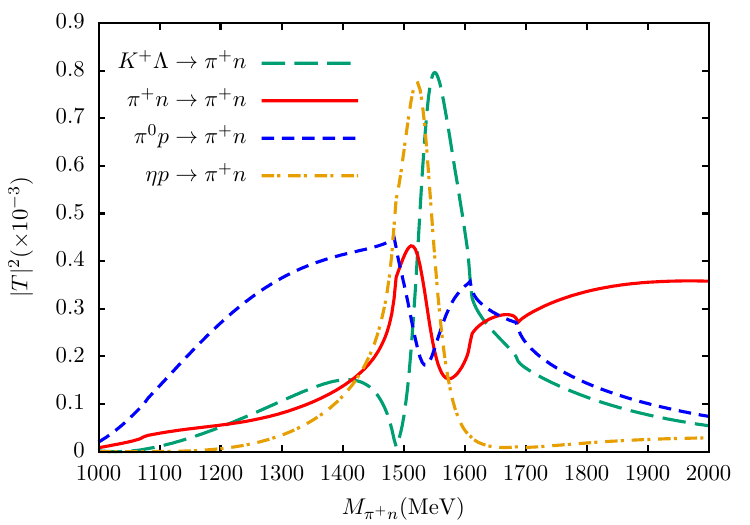}
	\caption{Modulus squared of the transition amplitudes $t_{MB} \to \pi^+n$ in S-wave.}
	\label{fig:pin-t}
\end{figure}

\begin{figure}[tbhp]
	\centering
	\includegraphics[scale=0.7]{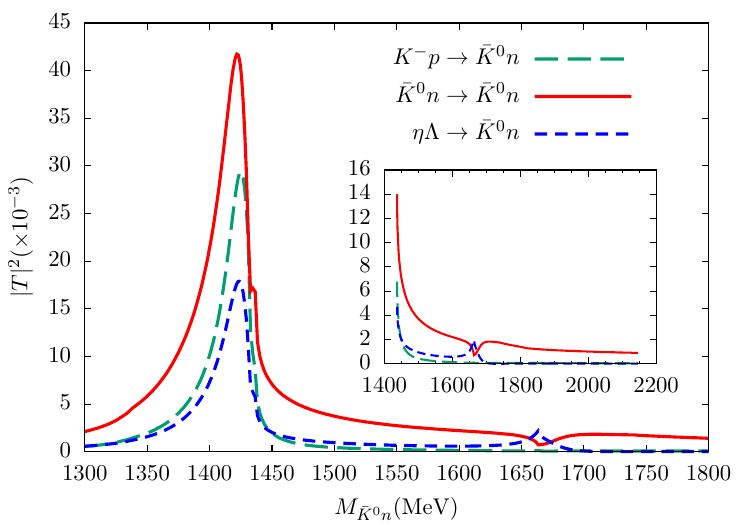}
	\caption{Modulus squared of the transition amplitudes $t_{M^{\prime}B^{\prime}} \to \bar{K}^0n$ in S-wave. The inner sub-figures show the results within the $\bar{K}^0n$ phase space.}
	\label{fig:kn-kn}
\end{figure}

We first examine the transition amplitudes $t_{MB \to \pi^+ n}$ and $t_{M'B' \to \bar{K}^0 n}$ that dynamically generate the resonances. Figure~\ref{fig:pin-t} shows $|t_{MB \to \pi^+ n}|^2$ for several initial channels. The green-dashed curve denotes the modulus squared of the transition amplitude $t_{K^+\Lambda \to \pi^+n}$, the red-solid curve denotes the modulus squared of the transition amplitude $t_{\pi^+n \to \pi^+n}$, the blue-dashed curve denotes the modulus squared of the transition amplitude $t_{\pi^0p \to \pi^+n}$, and the yellow-dashed-dotted curve denotes the modulus squared of the transition amplitude $t_{\eta p \to \pi^+n}$. A clear peak around 1500~MeV is visible, corresponding to the $N(1535)$ resonance. The $\pi^+ n \to \pi^+ n$ amplitude (red solid curve) shows the strong signal, as expected for the dominant decay channel.

Figure~\ref{fig:kn-kn} displays $|t_{M'B' \to \bar{K}^0 n}|^2$. The green-dashed curve denotes the modulus squared of the transition amplitude $t_{K^-p \to \bar{K}^0 n}$, the red-solid curve denotes the modulus squared of the transition amplitude $t_{\bar{K}^0 n \to \bar{K}^0 n}$, and the blue-dashed curve denotes the modulus squared of the transition amplitude $t_{\eta\Lambda \to \bar{K}^0 n}$. A prominent peak around 1420~MeV (the $\Lambda(1405)$) and a clear dip structure around 1670~MeV (the $\Lambda(1670)$) are seen in the $\bar{K}^0 n \to \bar{K}^0 n$ amplitude (red solid curve). The $\Lambda(1405)$ lies below the $\bar{K}^0 n$ threshold ($\sim 1430$~MeV) and thus contribute to the threshold of the $\bar{K}^0 n$ spectrum in this decay. The inset focuses on the physical region, highlighting the $\Lambda(1670)$ dip. The appearance of a dip (instead of a peak) is a characteristic prediction of the chiral unitary approach for $\Lambda(1670)$ in the $\bar{K}N$ channel and is consistent with its manifestation in $\bar{K}N \to \bar{K}N$ scattering data~\cite{Gopal:1976gs,Oset:2001cn}. This occurs due to a strong destructive interference between the resonant amplitude and the non-resonant background, driven by the phase motion of the $\Lambda(1670)$ pole located near the $\eta\Lambda$ threshold.

\subsection{Invariant Mass Distributions and Dalitz Plot for $\Lambda_c^+ \to \bar{K}^0 \pi^+ n$}
\begin{figure}[htbp]	
	\subfigure{
		\centering
		\includegraphics[scale=0.65]{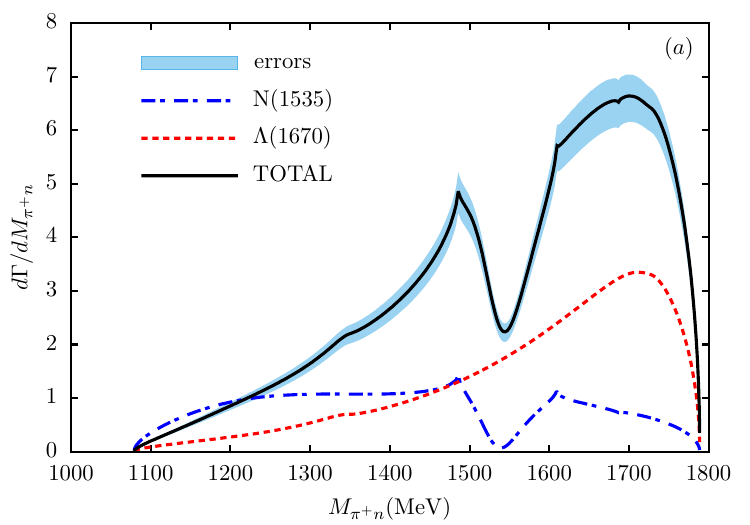}	
	}
	\subfigure{
		\centering
		\includegraphics[scale=0.65]{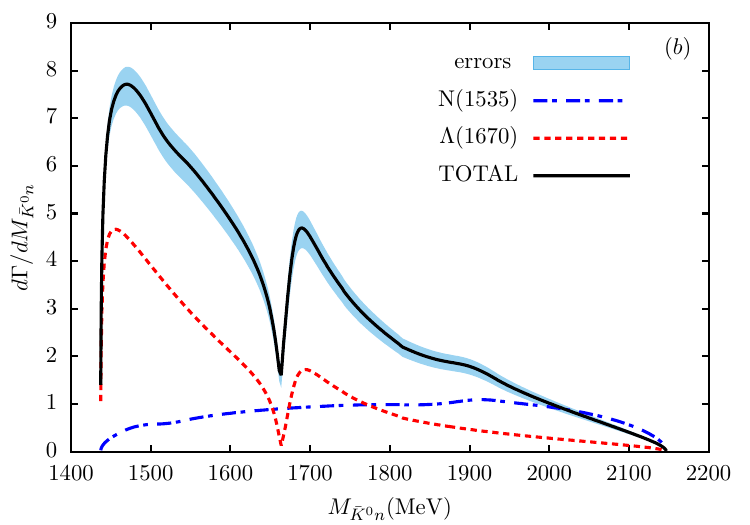}	
	}
	\caption{$\pi^+ n$ (a) and $\bar{K}^0 n$ (b) invariant mass distributions of the process $\Lambda_c^+ \to n \bar{K}^0\pi^+$. The error bands reflect the variation of the cutoff momentum parameter $|\tilde{q}_{\mathrm{max}}|$ within the range of 1100-1500~MeV on the linear features. All distributions are plotted in arbitrary units.}
	\label{fig:minv}
\end{figure}

\begin{figure}[tbhp]
	\centering
	\includegraphics[scale=0.8]{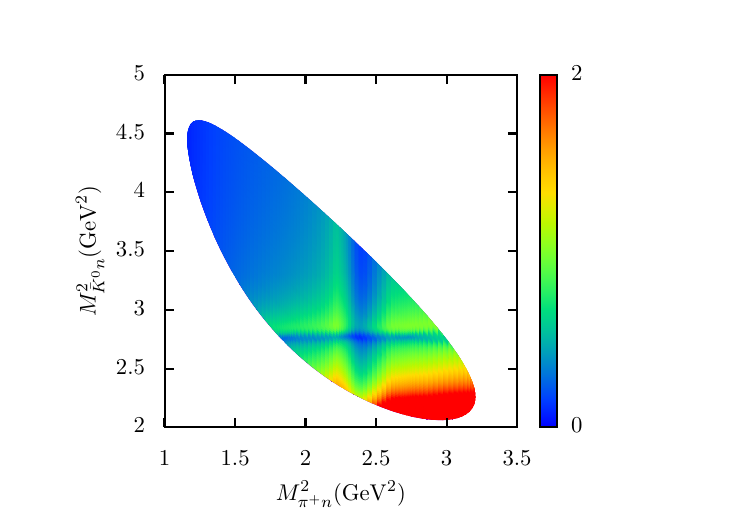}
	\caption{Dalitz plot for the process $\Lambda_c^+ \to n \bar{K}^0\pi^+$.}
	\label{fig:Dali}
\end{figure}

Our main results for the $\pi^+ n$ and $\bar{K}^0 n$ invariant mass distributions are presented in Fig.~\ref{fig:minv}. The individual contributions from $N(1535)$ (blue dash-dotted) and $\Lambda(1670)$ (red dotted), as well as their coherent sum with $\phi=0$ (black solid), are shown. Meanwhile, we have also considered the theoretical uncertainties from the cutoff momentum parameter $|\tilde{q}_{\mathrm{max}}|=1300\pm 200$~MeV in Eq.~(\ref{eq:gfunction_cutoff}), which are shown by the blue bands labeled as `errors' in Fig.~\ref{fig:minv}. One could find that the uncertainty of the parameter $|\tilde{q}_{\mathrm{max}}|=1300\pm 200$~MeV has little influence on the linear features near the resonance structure of the $\pi^+ n$ and $\bar{K}^0 n$ distributions.

In the $\pi^+ n$ distribution (Fig.~\ref{fig:minv}(a)), a pronounced narrow peak appears around 1500~MeV, stemming directly from the $N(1535)$ resonance.
The $\Lambda(1670)$ contribution to this distribution is relatively small and smooth. The peak's position and shape are robust predictions, which can be test in future experiments like BESIII, Belle II, or LHCb.

In the $\bar{K}^0 n$ distribution (Fig.~\ref{fig:minv}(b)), the most striking feature is a distinct dip around 1670~MeV, which is a signature of the $\Lambda(1670)$ resonance. The $N(1535)$ contribution provides a smooth background in this spectrum. The dip structure predicted here is qualitatively consistent with the behavior of $\Lambda(1670)$ observed in $\bar{K}N$ elastic scattering. This consistency supports the interpretation of $\Lambda(1670)$ as a state dynamically generated from coupled-channel interactions, whose observable line shape is highly process-dependent. In $\Lambda_c^+$ decay, the production mechanism and the interference with other amplitudes (e.g., from $N(1535)$ and non-resonant terms) can modify the depth and exact location of the dip, but the underlying mechanism producing the dip remains the same.

Our findings suggest that the $\Lambda(1670)$ contribution, which interferes with other isospin components, could play a key role in explaining the different effective isospin decompositions reported by BESIII and LHCb/Belle. A combined amplitude analysis of $\Lambda_c^+ \to p K^-\pi^+$ and $\Lambda_c^+ \to n K_S^0 \pi^+$, by taking into account these dynamically generated resonant contributions, would be crucial to resolve the apparent discrepancy, when the measurements of the process $\Lambda_c^+\to n \bar{K}^0 \pi^+$ are available.

The Dalitz plot for $\Lambda_c^+ \to \bar{K}^0 \pi^+ n$ in the plane of $M_{\pi^+ n}^2$ vs.\ $M_{\bar{K}^0 n}^2$ is shown in Fig.~\ref{fig:Dali}. A clear vertical band around $M_{\pi^+ n}^2 \approx 2.25$~GeV$^2$ corresponding to the $N(1535)$ signal and a horizontal band around $M_{\bar{K}^0 n}^2 \approx 2.79$ GeV$^2$ ($M_{\bar{K}^0 n} \approx 1670$~MeV) reflecting the $\Lambda(1670)$ dip structure, appears as a region of reduced event density. This plot provides a comprehensive visualization of how the two resonances populate the decay phase space.

\subsection{Effects of Relative Phase and Color Factor}

\begin{figure}[htbp]
	\centering		
	\subfigure{
		\centering
		\includegraphics[scale=0.65]{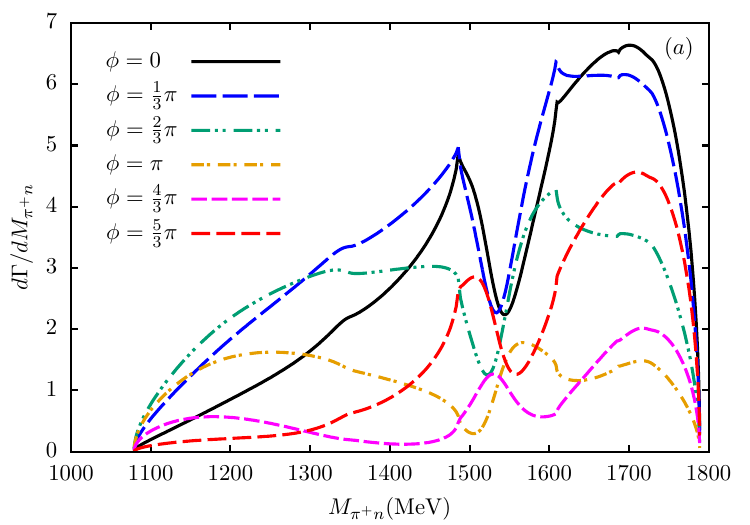}	
	}
	\subfigure{
		\centering
		\includegraphics[scale=0.65]{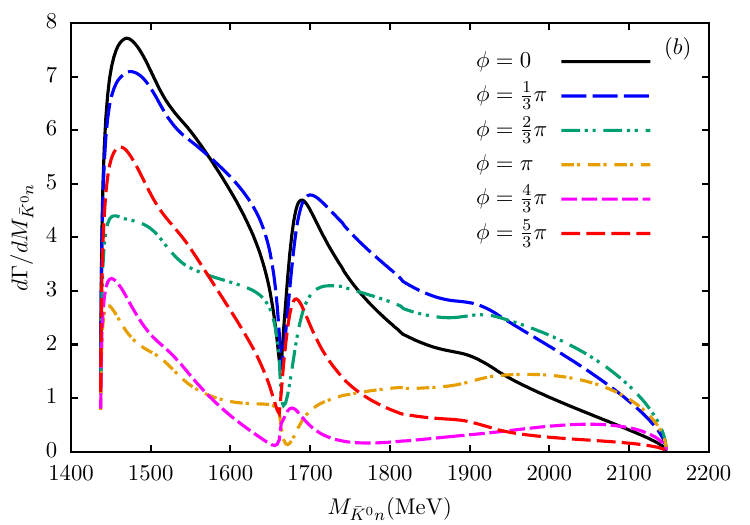}	
	}
	\caption{$\pi^+ n$ (a) and $\bar{K}^0 n$ (b) invariant mass distributions of the process $\Lambda_c^+ \to n \bar{K}^0\pi^+$ with the relative phase angle $\phi=0, \pi/3, 2\pi/3, \pi, 4\pi/3$, and $5\pi/3$, respectively. All distributions are plotted in arbitrary units.}
	\label{fig:minv-phi}
\end{figure}

\begin{figure}[htbp]
	\centering		
	\subfigure{
		\centering
		\includegraphics[scale=0.65]{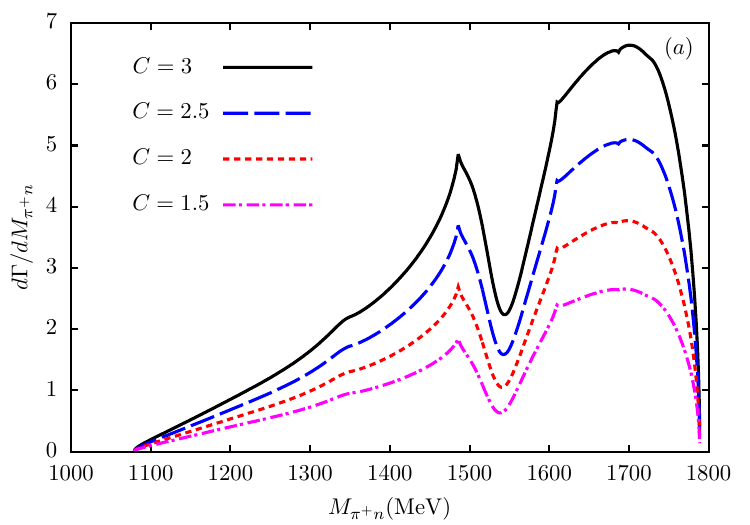}	
	}
	\subfigure{
		\centering
		\includegraphics[scale=0.65]{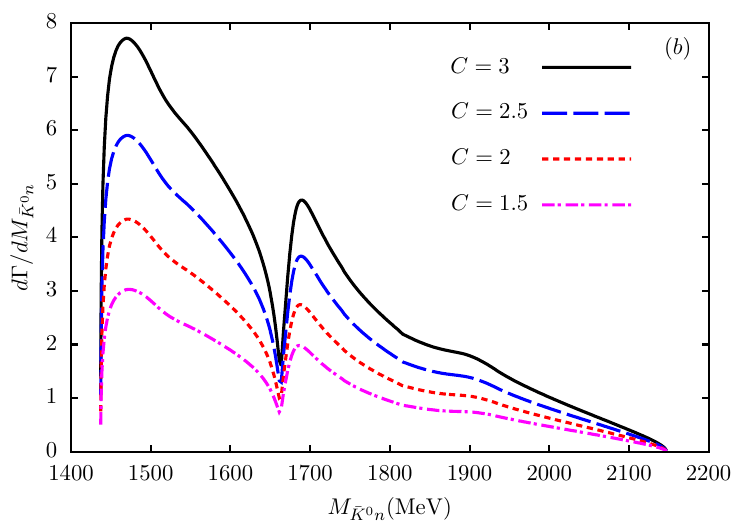}	
	}
	\caption{$\pi^+ n$ (a) and $\bar{K}^0 n$ (b) invariant mass distributions of the process $\Lambda_c^+ \to n \bar{K}^0\pi^+$ with the color factor $C=3, 2.5, 2$, and $1.5$, respectively. All distributions are plotted in arbitrary units.}
	\label{fig:minv-c}
\end{figure}

The coherent sum in Eq.~(\ref{Eq:M-total}) involves an unknown relative phase $\phi$ between the two production mechanisms. Figure~\ref{fig:minv-phi} shows the $\pi^+ n$ and $\bar{K}^0 n$ distributions for different values of $\phi$. The $N(1535)$ peak in the $\pi^+ n$ spectrum can be significantly enhanced or suppressed, due to interference with the $\Lambda(1670)$ amplitude. However, the position of the structure remains anchored near 1500~MeV. In the $\bar{K}^0 n$ spectrum, the $\Lambda(1670)$ dip remains a robust feature for all phase values, although its depth and the surrounding background vary. Future precise data could potentially constrain this phase.

We also investigate the sensitivity to the color factor $C$, which parametrizes the relative weight of the external emission diagram. Figure~\ref{fig:minv-c} shows that the variation of $C$ between 1.5 and 3 would change the overall normalization and the relative strength of the two resonant contributions, but does not qualitatively alter the peak and dip structures. The $N(1535)$ peak and the $\Lambda(1670)$ dip remain clearly identifiable.

\section{Summary} \label{sec:Conclusions}

In this work, we have presented a theoretical study of the Cabibbo-favored decay $\Lambda_c^+ \to n \bar{K}^0\pi^+$ within the chiral unitary approach, where the low-lying resonances $N(1535)$ and $\Lambda(1670)$ are dynamically generated from the coupled-channel interactions. The investigation is motivated by recent experimental observations: a large branching fraction for $\Lambda_c^+ \to n K_S^0 \pi^+$ that surpasses SU(3) predictions, and an apparent tension between the isospin decompositions of the $\bar{K}N$ system reported by different experiments.

Our calculations predict a clear narrow peak in the $\pi^+ n$ invariant mass distribution, associated with the $N(1535)$ resonance, and a distinct dip in the $\bar{K}^0 n$ distribution, associated with the $\Lambda(1670)$ resonance. The dip structure is qualitatively consistent with the known behavior of $\Lambda(1670)$ in $\bar{K}N$ scattering, supporting its interpretation as a dynamically generated state. The significant contributions of these resonances can naturally explain the enhanced branching fraction and may play a crucial role in reconciling the different experimental findings regarding the $\bar{K}N$ isospin composition.

The decay $\Lambda_c^+ \to n \bar{K}^0\pi^+$ thus emerges as a powerful process to probe the nature of two contentious low-lying baryon resonances: the $N(1535)$, whose internal structure (three-quark, pentaquark, or molecular) is debated, and the $\Lambda(1670)$, which exhibits fascinating process-dependent line shapes. The predicted distributions and Dalitz plot can be used to test our model. 

The absolute branching fraction of $\Lambda_c^+ \to n K_S^0 \pi^+$ has been measured to be at the percent level. We strongly encourage high-statistics studies of this decay channel by the BESIII, Belle II, LHCb, and the proposed Super Tau-Charm Factory Collaborations. Precise measurements of the $\pi^+ n$ and $\bar{K}^0 n$ invariant mass spectra, along with a full amplitude analysis, will be essential to confirm our predictions, constrain model parameters (like the relative phase $\phi$ and the $\Lambda(1670)$ subtraction constants), and ultimately shed light on the isospin dynamics and the internal structure of the $N(1535)$ and $\Lambda(1670)$ resonances.

\section*{Acknowledgments}
This work is supported by the National Key R$\&$D Program of China (No. 2024YFE0105200) and the National Natural Science Foundation of China under Grants No. 12475086, No. 12192263, No. 12305137 and No. 12205075.


\begin{thebibliography}{10}


\bibitem{BESIII:2025rda}
M.~Ablikim \textit{et al.} [BESIII],
Phys. Rev. D \textbf{112} (2025), 032006
doi:10.1103/csfm-p3h6
[arXiv:2506.02969 [hep-ex]].



\bibitem{Belle:2025voy}
I.~Adachi \textit{et al.} [Belle and Belle-II],
Phys. Rev. D \textbf{112} (2025) no.1, 012013
doi:10.1103/4mnw-tvks
[arXiv:2503.04371 [hep-ex]].

\bibitem{Wang:2024jyk}
E.~Wang, L.~S.~Geng, J.~J.~Wu, J.~J.~Xie and B.~S.~Zou,
Chin. Phys. Lett. \textbf{41} (2024) no.10, 101401
doi:10.1088/0256-307X/41/10/101401
[arXiv:2406.07839 [hep-ph]].

\bibitem{Feng:2020jvp}
X.~C.~Feng, L.~L.~Wei, M.~Y.~Duan, E.~Wang and D.~M.~Li,
Phys. Lett. B \textbf{846} (2023), 138185
doi:10.1016/j.physletb.2023.138185
[arXiv:2009.08600 [hep-ph]].

\bibitem{Zeng:2020och}
C.~H.~Zeng, J.~X.~Lu, E.~Wang, J.~J.~Xie and L.~S.~Geng,
Phys. Rev. D \textbf{102} (2020) no.7, 076009
doi:10.1103/PhysRevD.102.076009
[arXiv:2006.15547 [hep-ph]].

\bibitem{Wang:2020pem}
Z.~Wang, Y.~Y.~Wang, E.~Wang, D.~M.~Li and J.~J.~Xie,
Eur. Phys. J. C \textbf{80} (2020) no.9, 842
doi:10.1140/epjc/s10052-020-8347-2
[arXiv:2004.01438 [hep-ph]].


\bibitem{BESIII:2016yrc}
M.~Ablikim \textit{et al.} [BESIII],
Phys. Rev. Lett. \textbf{118} (2017) no.11, 112001
doi:10.1103/PhysRevLett.118.112001
[arXiv:1611.02797 [hep-ex]].


\bibitem{BESIII:2023pia}
M.~Ablikim \textit{et al.} [BESIII],
Phys. Rev. D \textbf{109} (2024) no.7, 072010
doi:10.1103/PhysRevD.109.072010
[arXiv:2311.17131 [hep-ex]].



\bibitem{Geng:2018upx}
C.~Q.~Geng, Y.~K.~Hsiao, C.~W.~Liu and T.~H.~Tsai,
Phys. Rev. D \textbf{99} (2019) no.7, 073003
doi:10.1103/PhysRevD.99.073003
[arXiv:1810.01079 [hep-ph]].

\bibitem{Cen:2019ims}
J.~Y.~Cen, C.~Q.~Geng, C.~W.~Liu and T.~H.~Tsai,
Eur. Phys. J. C \textbf{79} (2019) no.11, 946
doi:10.1140/epjc/s10052-019-7467-z
[arXiv:1906.01848 [hep-ph]].

\bibitem{LHCb:2022sck}
R.~Aaij \textit{et al.} [LHCb],
Phys. Rev. D \textbf{108} (2023) no.1, 012023
doi:10.1103/PhysRevD.108.012023
[arXiv:2208.03262 [hep-ex]].

\bibitem{Belle:2022cbs}
S.~B.~Yang \textit{et al.} [Belle],
Phys. Rev. D \textbf{108} (2023) no.3, L031104
doi:10.1103/PhysRevD.108.L031104
[arXiv:2209.00050 [hep-ex]].

\bibitem{Zhang:2024jby}
S.~C.~Zhang, M.~Y.~Duan, W.~T.~Lyu, G.~Y.~Wang, J.~Y.~Zhu and E.~Wang,
Eur. Phys. J. C \textbf{84} (2024) no.12, 1253
doi:10.1140/epjc/s10052-024-13616-6
[arXiv:2405.14235 [hep-ph]].

\bibitem{Duan:2024okk}
M.~Y.~Duan, M.~Bayar and E.~Oset,
Phys. Lett. B \textbf{857} (2024), 139003
doi:10.1016/j.physletb.2024.139003
[arXiv:2407.01410 [hep-ph]].

\bibitem{Lu:2016ogy}
C.~D.~L{\"u}, W.~Wang and F.~S.~Yu,
Phys. Rev. D \textbf{93} (2016) no.5, 056008
doi:10.1103/PhysRevD.93.056008
[arXiv:1601.04241 [hep-ph]].



\bibitem{Liu:2005pm}
B.~C.~Liu and B.~S.~Zou,
Phys. Rev. Lett. \textbf{96} (2006) no.4, 042002
doi:10.1103/PhysRevLett.96.042002
[arXiv:nucl-th/0503069 [nucl-th]].

\bibitem{Geng:2008cv}
L.~S.~Geng, E.~Oset, B.~S.~Zou and M.~Doring,
Phys. Rev. C \textbf{79} (2009), 025203
doi:10.1103/PhysRevC.79.025203
[arXiv:0807.2913 [hep-ph]].


\bibitem{Hannelius:2000gu}
L.~Hannelius and D.~O.~Riska,
Phys. Rev. C \textbf{62} (2000), 045204
doi:10.1103/PhysRevC.62.045204
[arXiv:hep-ph/0001325 [hep-ph]].

\bibitem{Helminen:2000jb}
C.~Helminen and D.~O.~Riska,
Nucl. Phys. A \textbf{699} (2002), 624-648
doi:10.1016/S0375-9474(01)01294-5
[arXiv:nucl-th/0011071 [nucl-th]].

\bibitem{Zhang:2004xt}
A.~Zhang, Y.~R.~Liu, P.~Z.~Huang, W.~Z.~Deng, X.~L.~Chen and S.~L.~Zhu,
HEPNP \textbf{29} (2005), 250

\bibitem{Zou:2007mk}
B.~S.~Zou,
Eur. Phys. J. A \textbf{35} (2008), 325-328
doi:10.1140/epja/i2007-10561-8
[arXiv:0711.4860 [nucl-th]].

\bibitem{Oset:1997it}
E.~Oset and A.~Ramos,
Nucl. Phys. A \textbf{635} (1998), 99-120
doi:10.1016/S0375-9474(98)00170-5
[arXiv:nucl-th/9711022 [nucl-th]].

\bibitem{Jido:2003cb}
D.~Jido, J.~A.~Oller, E.~Oset, A.~Ramos and U.~G.~Meissner,
Nucl. Phys. A \textbf{725} (2003), 181-200
doi:10.1016/S0375-9474(03)01598-7
[arXiv:nucl-th/0303062 [nucl-th]].

\bibitem{Kaiser:1996js}
N.~Kaiser, T.~Waas and W.~Weise,
Nucl. Phys. A \textbf{612} (1997), 297-320
doi:10.1016/S0375-9474(96)00321-1
[arXiv:hep-ph/9607459 [hep-ph]].

\bibitem{Nieves:2001wt}
J.~Nieves and E.~Ruiz Arriola,
Phys. Rev. D \textbf{64} (2001), 116008
doi:10.1103/PhysRevD.64.116008
[arXiv:hep-ph/0104307 [hep-ph]].

\bibitem{Inoue:2001ip}
T.~Inoue, E.~Oset and M.~J.~Vicente Vacas,
Phys. Rev. C \textbf{65} (2002), 035204
doi:10.1103/PhysRevC.65.035204
[arXiv:hep-ph/0110333 [hep-ph]].

\bibitem{Doring:2008sv}
M.~Doring, E.~Oset and B.~S.~Zou,
Phys. Rev. C \textbf{78} (2008), 025207
doi:10.1103/PhysRevC.78.025207
[arXiv:0805.1799 [nucl-th]].

\bibitem{Wang:2015pcn}
E.~Wang, H.~X.~Chen, L.~S.~Geng, D.~M.~Li and E.~Oset,
Phys. Rev. D \textbf{93} (2016) no.9, 094001
doi:10.1103/PhysRevD.93.094001
[arXiv:1512.01959 [hep-ph]].

\bibitem{Lu:2016roh}
J.~X.~Lu, E.~Wang, J.~J.~Xie, L.~S.~Geng and E.~Oset,
Phys. Rev. D \textbf{93} (2016), 094009
doi:10.1103/PhysRevD.93.094009
[arXiv:1601.00075 [hep-ph]].

\bibitem{Xie:2017erh}
J.~J.~Xie and L.~S.~Geng,
Phys. Rev. D \textbf{96} (2017) no.5, 054009
doi:10.1103/PhysRevD.96.054009
[arXiv:1704.05714 [hep-ph]].

\bibitem{Pavao:2018wdf}
R.~Pavao, S.~Sakai and E.~Oset,
Phys. Rev. C \textbf{98} (2018) no.1, 015201
doi:10.1103/PhysRevC.98.015201
[arXiv:1802.07882 [nucl-th]].

\bibitem{Lyu:2023aqn}
W.~T.~Lyu, Y.~H.~Lyu, M.~Y.~Duan, G.~Y.~Wang, D.~Y.~Chen and E.~Wang,
Eur. Phys. J. C \textbf{85} (2025) no.2, 123
doi:10.1140/epjc/s10052-025-13805-x
[arXiv:2310.11139 [hep-ph]].

\bibitem{Li:2024rqb}
Y.~Li, S.~W.~Liu, E.~Wang, D.~M.~Li, L.~S.~Geng and J.~J.~Xie,
Phys. Rev. D \textbf{110} (2024) no.7, 074010
doi:10.1103/PhysRevD.110.074010
[arXiv:2406.01209 [hep-ph]].

\bibitem{Li:2025gvo}
M.~Y.~Li, W.~T.~Lyu, L.~J.~Liu and E.~Wang,
Phys. Rev. D \textbf{111} (2025) no.3, 034046
doi:10.1103/PhysRevD.111.034046
[arXiv:2501.02859 [hep-ph]].

\bibitem{Song:2025eko}
J.~Song, M.~Bayar, Y.~Y.~Li and E.~Oset,
Eur. Phys. J. C \textbf{85} (2025) no.10, 1114
doi:10.1140/epjc/s10052-025-14870-y
[arXiv:2507.19240 [hep-ph]].

\bibitem{Li:2026lbo}
Y.~Li, E.~Wang, L.~S.~Geng and J.~J.~Xie,
Phys. Rev. D \textbf{113} (2026) no.5, 054039
doi:10.1103/7tbd-7krm
[arXiv:2601.13668 [hep-ph]].


\bibitem{Liu:2015ktc}
Z.~W.~Liu, W.~Kamleh, D.~B.~Leinweber, F.~M.~Stokes, A.~W.~Thomas and J.~J.~Wu,
Phys. Rev. Lett. \textbf{116} (2016) no.8, 082004
doi:10.1103/PhysRevLett.116.082004
[arXiv:1512.00140 [hep-lat]].

\bibitem{Guo:2022hud}
D.~Guo and Z.~W.~Liu,
Phys. Rev. D \textbf{105} (2022) no.11, 114039
doi:10.1103/PhysRevD.105.114039
[arXiv:2201.11555 [hep-ph]].

\bibitem{Abell:2023nex}
C.~D.~Abell, D.~B.~Leinweber, Z.~W.~Liu, A.~W.~Thomas and J.~J.~Wu,
Phys. Rev. D \textbf{108} (2023) no.9, 094519
doi:10.1103/PhysRevD.108.094519
[arXiv:2306.00337 [hep-lat]].

\bibitem{Molina:2023jov}
R.~Molina, C.~W.~Xiao, W.~H.~Liang and E.~Oset,
Phys. Rev. D \textbf{109} (2024) no.5, 054002
doi:10.1103/PhysRevD.109.054002
[arXiv:2310.12593 [hep-ph]].

\bibitem{Li:2023pjx}
H.~P.~Li, J.~Song, W.~H.~Liang, R.~Molina and E.~Oset,
Eur. Phys. J. C \textbf{84} (2024) no.7, 656
doi:10.1140/epjc/s10052-024-13015-x
[arXiv:2311.14365 [hep-ph]].



\bibitem{Belle:2020xku}
J.~Y.~Lee \textit{et al.} [Belle],
Phys. Rev. D \textbf{103} (2021) no.5, 052005
doi:10.1103/PhysRevD.103.052005
[arXiv:2008.11575 [hep-ex]].



\bibitem{Gopal:1976gs}
G.~P.~Gopal \textit{et al.} [Rutherford-London],
Nucl. Phys. B \textbf{119} (1977), 362-400
doi:10.1016/0550-3213(77)90002-5

\bibitem{Oset:2001cn}
E.~Oset, A.~Ramos and C.~Bennhold,
Phys. Lett. B \textbf{527} (2002), 99-105
[erratum: Phys. Lett. B \textbf{530} (2002), 260-260]
doi:10.1016/S0370-2693(01)01523-4
[arXiv:nucl-th/0109006 [nucl-th]].

\bibitem{CrystalBall:2001uhc}
A.~Starostin \textit{et al.} [Crystal Ball],
Phys. Rev. C \textbf{64} (2001), 055205
doi:10.1103/PhysRevC.64.055205

\bibitem{Zhong:2008km}
X.~H.~Zhong and Q.~Zhao,
Phys. Rev. C \textbf{79} (2009), 045202
doi:10.1103/PhysRevC.79.045202
[arXiv:0811.4212 [nucl-th]].

\bibitem{Liu:2023xvy}
J.~J.~Liu, Z.~W.~Liu, K.~Chen, D.~Guo, D.~B.~Leinweber, X.~Liu and A.~W.~Thomas,
Phys. Rev. D \textbf{109} (2024) no.5, 054025
doi:10.1103/PhysRevD.109.054025
[arXiv:2312.13072 [hep-ph]].

\bibitem{Xie:2016evi}
J.~J.~Xie and L.~S.~Geng,
Eur. Phys. J. C \textbf{76} (2016) no.9, 496
doi:10.1140/epjc/s10052-016-4342-z
[arXiv:1604.02756 [nucl-th]].

\bibitem{Wang:2022nac}
G.~Y.~Wang, N.~C.~Wei, H.~M.~Yang, E.~Wang, L.~S.~Geng and J.~J.~Xie,
Phys. Rev. D \textbf{106} (2022) no.5, 056001
doi:10.1103/PhysRevD.106.056001
[arXiv:2206.01425 [hep-ph]].

\bibitem{Miyahara:2015cja}
K.~Miyahara, T.~Hyodo and E.~Oset,
Phys. Rev. C \textbf{92} (2015) no.5, 055204
doi:10.1103/PhysRevC.92.055204
[arXiv:1508.04882 [nucl-th]].



\bibitem{Duan:2024czu}
M.~Y.~Duan, W.~T.~Lyu, C.~W.~Xiao, E.~Wang, J.~J.~Xie, D.~Y.~Chen and E.~Oset,
Phys. Rev. D \textbf{111} (2025) no.1, 016004
doi:10.1103/PhysRevD.111.016004
[arXiv:2410.16078 [hep-ph]].

\bibitem{Lyu:2024qgc}
W.~T.~Lyu, S.~C.~Zhang, G.~Y.~Wang, J.~J.~Wu, E.~Wang, L.~S.~Geng and J.~J.~Xie,
Phys. Rev. D \textbf{110} (2024) no.5, 054020
doi:10.1103/PhysRevD.110.054020
[arXiv:2405.09226 [hep-ph]].

\bibitem{Wang:2022xqc}
Z.~Y.~Wang, S.~Q.~Luo, Z.~F.~Sun, C.~W.~Xiao and X.~Liu,
Phys. Rev. D \textbf{106} (2022) no.9, 096026
doi:10.1103/PhysRevD.106.096026
[arXiv:2211.08946 [hep-ph]].


\bibitem{Li:2025exm}
Y.~Li, W.~T.~Lyu, G.~Y.~Wang, L.~Li, W.~C.~Yan and E.~Wang,
Phys. Rev. D \textbf{111} (2025) no.5, 054011
doi:10.1103/PhysRevD.111.054011
[arXiv:2501.14385 [hep-ph]].

\bibitem{Zhang:2026igc}
S.~C.~Zhang, W.~T.~Lyu, G.~Y.~Wang, B.~Q.~Ma and E.~Wang,
Chin. Phys. Lett. \textbf{43} (2026) no.5, 050203
doi:10.1088/0256-307X/43/5/050203
[arXiv:2601.12778 [hep-ph]].


\bibitem{Miyahara:2016yyh}
K.~Miyahara, T.~Hyodo, M.~Oka, J.~Nieves and E.~Oset,
Phys. Rev. C \textbf{95} (2017) no.3, 035212
doi:10.1103/PhysRevC.95.035212
[arXiv:1609.00895 [nucl-th]].

\bibitem{Pavao:2017cpt}
R.~P.~Pavao, W.~H.~Liang, J.~Nieves and E.~Oset,
Eur. Phys. J. C \textbf{77} (2017) no.4, 265
doi:10.1140/epjc/s10052-017-4836-3
[arXiv:1701.06914 [hep-ph]].


\bibitem{Bramon:1992kr}
A.~Bramon, A.~Grau and G.~Pancheri,
Phys. Lett. B \textbf{283} (1992), 416-420
doi:10.1016/0370-2693(92)90041-2

\bibitem{Oset:2016lyh}
E.~Oset, W.~H.~Liang, M.~Bayar, J.~J.~Xie, L.~R.~Dai, M.~Albaladejo, M.~Nielsen, T.~Sekihara, F.~Navarra and L.~Roca, \textit{et al.}
Int. J. Mod. Phys. E \textbf{25} (2016), 1630001
doi:10.1142/S0218301316300010
[arXiv:1601.03972 [hep-ph]].


\bibitem{Hyodo:2008xr}
T.~Hyodo, D.~Jido and A.~Hosaka,
Phys. Rev. C \textbf{78} (2008), 025203
doi:10.1103/PhysRevC.78.025203
[arXiv:0803.2550 [nucl-th]].


\bibitem{Doring:2005bx}
M.~Doring, E.~Oset and D.~Strottman,
Phys. Rev. C \textbf{73} (2006), 045209
doi:10.1103/PhysRevC.73.045209
[arXiv:nucl-th/0510015 [nucl-th]].

\bibitem{ParticleDataGroup:2024cfk}
S.~Navas \textit{et al.} [Particle Data Group],
Phys. Rev. D \textbf{110} (2024) no.3, 030001
doi:10.1103/PhysRevD.110.030001










\end{thebibliography}

\end{document}